\documentclass[11pt,a4paper]{article}

\usepackage{a4wide}
\usepackage{amsmath}
\usepackage{bm}
\usepackage{amssymb}
\usepackage{hyperref}
\usepackage{epsfig}
\usepackage{epic}
\usepackage{mathrsfs}

\newcommand{\p}{\partial}

\newcommand{\Tr}{{\rm Tr \,}}
\newcommand{\Op}{\mathcal{O}}
\newcommand{\fldZ}{\mathcal{Z}}

\newcommand{\fldD}{\mathcal{D}}

\newcommand{\alg}[1]{\mathfrak{#1}}
\newcommand{\superN}{\mathcal{N}}
\newcommand{\gym}{g\indups{YM}}
\newcommand{\indups}[1]{_{\mathrm{\scriptscriptstyle #1}}}
\newcommand{\sfrac}[2]{{\textstyle\frac{#1}{#2}}}

\newcommand{\I}{\mathrm{I}}
\newcommand{\II}{\mathrm{II}}
\newcommand{\III}{\mathrm{III}}
\newcommand{\IV}{\mathrm{IV}}

\makeatletter
\def\mr@ignsp#1 {\ifx\:#1\@empty\else #1\expandafter\mr@ignsp\fi}%
\newcommand{\multiref}[1]{\begingroup%\let\protect\string%
\xdef\mr@no@sparg{\expandafter\mr@ignsp#1 \: }%
\def\mr@comma{}%
\@for\mr@refs:=\mr@no@sparg\do{\mr@comma\def\mr@comma{,}\ref{\mr@refs}}%
\endgroup}
\makeatother
\long\def\symbolfootnote[#1]#2{\begingroup%
\def\thefootnote{\fnsymbol{footnote}}\footnote[#1]{#2}\endgroup}

\newcommand{\Appref}[1]{Appendix~\multiref{#1}}
\newcommand{\appref}[1]{App.~\multiref{#1}}

\numberwithin{equation}{section}

\begin{document}

%%%%%%%%%%%%%%%%%%%%%%%%%%%%%%%%%%%%%%%%%%%%%%%%%%%%%%%%%%%%%%%%%%%%%%
\thispagestyle{empty}

\begin{flushright}\footnotesize
%\texttt{hep-th/yymmnnn}\\
\texttt{AEI-2008-082}\\
\vspace{0.5cm}
\end{flushright}
\setcounter{footnote}{0}

\begin{center}
{\Large\textbf{\mathversion{bold} Analytic three-loop Solutions for $\superN=4$ SYM
			          Twist Operators }}
\vspace{15mm}

{\sc Anatoly~V.~Kotikov$^a$, Adam Rej$^{b,}$\symbolfootnote[1]{Current address: \textit{The Blackett Laboratory, Imperial College, London SW7 2AZ, U.K.}} and Stefan Zieme$^b$}\\[10mm]

{\it $^a$ Bogoliubov Laboratory of Theoretical Physics\\
    Joint Institute for Nuclear Research\\
    141980 Dubna, Russia}\\[10mm]

{\it $^b$ Max-Planck-Institut f\"ur Gravitationsphysik\\
    Albert-Einstein-Institut \\
    Am M\"uhlenberg 1, D-14476 Potsdam, Germany}\\[7mm]

\texttt{kotikov@theor.jinr.ru}\\
\texttt{arej@aei.mpg.de}\\
\texttt{stzieme@aei.mpg.de}\\[30mm]

\textbf{Abstract}\\[2mm]
\end{center}

\noindent{We introduce a method to obtain the analytic solution of the
higher-order Baxter equation for twist-two and twist-three operators of planar $\superN=4$
SYM. Our result proofs the conjectured formula for the three-loop anomalous dimension
of twist-two operators. As such we derive the maximally transcendental
part of the corresponding three-loop QCD result from the maximal
supersymmetric gauge theory in four dimension purely by methods of
integrability.}

%%%%%%%%%%%%%%%%%%%%%%%%%%%%%%%%%%%%%%%%%%%%%%%%%%%%%%%%%%%%%%%%%%%%%%
\newpage

\setcounter{page}{1}
%%%%%%%%%%%%%%%%%%%%%%%%%%%%%%%%%%%%%%%%%%%%%%%%%%%%%%%%%%%%%%%%%%%%%%

%%%%%%%%%%%%%%%%%%%%%%%%%%%%%%%%%%%%%%%%%%%%%%%%%%%%%%%%%%%%%%%%%%%%%%
\section{Introduction}

The exploration of the AdS/CFT correspondence using the methods of
integrability has led to a first insight into its interpolating
property between weak and strong coupling. Since the discovery of
integrable structures on both sides of the correspondence
\cite{Integrable Structures} many techniques have been introduced and
developed which will hopefully lead to the complete solution of the planar
spectral problem in a finite volume.

Integrable structures in four-dimensional quantum field theories are not new
and have also been found in non-supersymmetric theories before, see for example
\cite{Lipatov:1993yb,Lipatov:1994xy,Faddeev Korchemsky} and \cite{Integrability in QCD}
and references therein. However, superconformal invariance preserves integrability of the
maximal supersymmetric planar $\superN=4$ SYM to higher, possibly all, orders.
The factorized scattering of elementary excitations of a super-spin chain
\cite{super spin chain} describing $\superN=4$ SYM at one-loop is extended
to higher-loop order \cite{Factorized S-matrix} and allows the determination
of the asymptotic spectrum in terms of long-range Bethe ansatz equations
\cite{LongRange}. The ambiguity of a phase factor of the S-matrix, determined by
symmetry solely \cite{su(2|2) s-matrix}, is also fixed by a constraining
crossing-symmetry \cite{Janik} supplemented by an unequivocally specification
of its solution \cite{BHL,BES}.

The operators which gained the most attention so far are
twist operators with lowest possible anomalous dimension, whose thermodynamical
behavior is governed by the so-called
universal scaling function \cite{ES,BES}. It was shown that the strong coupling
expansion of this scaling function, see \cite{Benna:2006nd} and \cite{StrongCoupling}
and references therein, correctly reproduces
the know string results \cite{Cusp string results} of the corresponding dual
state to two-loop order in perturbation theory.

To this class of operators belong the shortest possible local composite operators of
the $\superN=4$ theory, the twist-two operators. For a simple
representative of these one starts from the protected half-BPS states
$\Tr \fldZ^2$ and inserts $M$ covariant derivatives $\fldD$
\begin{equation}\label{twisttwo}
\Tr \left( \fldZ\, \fldD^M \, \fldZ\,\right) + \ldots\, .
\end{equation}

In the spin chain picture this is a non-compact $\alg{sl}(2)$
spin $=-\sfrac{1}{2}$ length-two Heisenberg magnet with $M$ magnons.
The dots indicate the mixing of all states where the covariant
derivatives may act on any of the two fields. For each even $M$
there is precisely one highest weight non-BPS state whose total scaling dimension is
\begin{equation}\label{dimension}
\Delta=2+M+\gamma(g,M) \,, \qquad {\rm with} \qquad
\gamma(g,M)=\sum_{\ell=1}^\infty  \gamma^{}_{2\ell}(M)\,g^{2\ell}\, ,
\end{equation}
where $\gamma(g,M)$ is the anomalous part of the dimension that
depends on the coupling constant $g^2=\sfrac{\lambda}{16\,\pi^2}$
and $\lambda=N\, \gym^2$ is the 't Hooft coupling constant.

For twist-two closed expressions for the anomalous part $\gamma(g,M)$ of the dimension are known
to two-loop order from explicit field-theory calculations \cite{Kotikov:2003fb},
and at three-loops from a solid conjecture
\cite{Kotikov:2004er} extracted by the principle of maximum
transcendentality \cite{Kotikov:2002ab} from the QCD splitting
functions at three-loops \cite{Moch:2004pa}. To this order
the anomalous part $\gamma(g,M)$ of the dimension can also be reliably
computed by the asymptotic Bethe ansatz \cite{Factorized S-matrix} for fixed
values of $M$. Up to relatively high values of $M$ it was checked
\cite{Factorized S-matrix} that it coincides with the two- and three-loop
anomalous dimensions of the twist-two operators, which are known in terms of
nested harmonic sums as obtained in \cite{Kotikov:2003fb,Kotikov:2004er}.
Due to wrapping effects that appear at four-loop order,
these operators have also been used to show the incorrectness of
the asymptotic Bethe ansatz at this order of perturbation theory
\cite{Dressing&Wrapping}.

It had not been known how to obtain these closed results in
a rigorous analytical way from the Bethe ansatz. This problem will finally be solved
in the present paper. By assuming the principle of maximum
transcendentality it is possible to make a general ansatz that contains harmonic sums
of the proper degree with unknown coefficients and to fit these with a high numerical
precision from the Bethe ansatz, but this principle is not granted in general for
other natural operators \cite{Nestin&Dressing}.

Therefore, we introduce a systematic approach to derive these results from
the Baxter equation that can be obtained from the $\alg{sl}(2)$
Bethe ansatz. As such we derive the maximally transcendental part
of the corresponding QCD computation from the maximal supersymmetric
Yang-Mills theory in four dimensions solely by means of integrability.
As an application of our method we also derive the three-loop anomalous
dimension of twist-three operators.

The method presented in this paper is based on deforming the one-loop solution.
The precise structure of this deformations can either be obtained by taking the
Mellin transformation of the perturbative Baxter equation or simply by an educated
guess, based on comparing the higher-loop Baxter equation with the difference equation
of a general deformation of the one-loop solution. In the main text we follow the latter,
while the aspects of the solution in Mellin space are given in the appendix.

We begin our analysis by introducing the perturbative Baxter equation, which has
been proposed in \cite{BelitskyBaxter}. After constructing the solution for twist-two
operators in an pedagogical way, we will apply the same techniques to twist-three
operators but only state the final result. Lengthy and detailed computations are
shifted to the Appendix as well as the complete calculation of anomalous
dimension. We close with a summary and give some ideas of further applications.

%%%%%%%%%%%%%%%%%%%%%%%%%%%%%%%%%%%%%%%%%%%%%%%%%%%%%%%%%%%%%%%%%%%%%%
\section{Baxter equation}

The Bethe roots that parametrize a solution of the Bethe ansatz equation
obey the following expansion in the coupling constant $g^2$ up to a loop
order $\ell$
\begin{equation}
u_k(g^2)=\sum_{i=0}^{\ell-1} \, g^{2i} \, u_k^{(i)} \,.
\end{equation}
They are given by zeros of the corresponding Baxter function $Q(u)$ which
also exhibits a similar expansion
\begin{equation}\label{Q expansion}
Q(u)=\prod_{k=1}^{M}\big(u-u_k(g^2)\big)=\sum_{i=0}^{\ell-1} \, g^{2i} \, Q^{(i)}\,.
\end{equation}
To three-loop-order $Q(u)$ satisfies the Baxter equation in form of
a second-order finite difference equation \cite{BelitskyBaxter}
\begin{equation}
\Delta_+(x^{+})\,Q(u+i)+\Delta_-(x^{-})\,Q(u-i)=t_L(x)\,Q(u)\,.
\end{equation}
The $x^{\pm}$ variables are defined through the spectral parameter $u$ in the following way \cite{BDS}
\begin{equation}
x^{\pm}(u)=\frac{u}{2}\left(1+\sqrt{1-\frac{4g^2}{u^2}}\right)\,.
\end{equation}
To the first three orders of perturbation theory, $\ell=2$, the functions
$\Delta_\pm$ have the following form
\begin{equation}\label{eq:GAMMA}
\Delta_\pm(x)=x^L \, \exp\left( \frac{ig^2}{x} \gamma^{(0)}_\pm
              +\frac{ig^4}{x}\gamma^{(1)}_\pm
              -\frac{g^4}{x^2}\frac{d^2}{du^2}
	       \left.\log Q^{(0)}(u)\right|_{u=\pm \tfrac{i}{2}}\right)\,.
\end{equation}
The auxiliary transfer matrix $t_L(x)$ for twist-two and three operators
is given by\footnote{Note, that $q_3$ is zero for the
twist-three ground state \cite{Dressing&Wrapping}.}
\begin{equation}
t_{L}(x)=2 u^L + q_2(g^2) \, u^{L-2}\,, \quad L=2,3 \,,
\end{equation}
where the charge $q_2$ is also to be expanded in $g^2$. Using \eqref{Q expansion}
one can read off the coefficient functions $q_2^{(i)}$, see below.
The resulting expressions for the anomalous dimensions are given by
\begin{equation}
\gamma(g^2)=\gamma_+(g^2)-\gamma_-(g^2)
\end{equation}
and
\begin{equation}\label{eq:gamma}
\gamma_\pm(g^2)=\left(2\,g^2 \frac{d}{du}+g^4\frac{d^3}{du^3}
		+\frac{g^6}{6}\frac{d^5}{du^5} \right)
                \big(i \log Q(u) \big)\Big|_{u=\pm\frac{i}{2}}
               =\sum_{i=0}^{\ell-1} \, g^{2(i+1)} \, \gamma_\pm^{(i)}\,.
\end{equation}

%
%%%%%%%%%%%%%%%%%%%%%%%%%%%%%%%%%%%%%%%%%%%%%%%%%%%%%%%%%%%%%%%%%%%%%%
\section{Twist-two operators}\label{Twist-two operators}

We will now explain how to find the higher order Baxter function for
operators that have been analyzed to a large extend in AdS/CFT, twist-two
operators. After restating the one-loop problem with its known solution
which for completeness is also given in \Appref{App:Mellin}, we focus on the explicit
construction of the higher loop solution. The main idea of the construction is based on deforming
the one-loop solution in a suitable manner. As mentioned before, one can also perform
the complete computation in Mellin space and verify the result, as included in \Appref{App:Mellin}.

\medskip

The leading order Baxter equation is given by
\begin{equation}\label{Baxter LO}
\left(u+\tfrac{i}{2} \right)^2 Q^{(0)}(u+i)+\left(u-\tfrac{i}{2}\right)^2 Q^{(0)}(u-i)
-t^{(0)}_2(u)Q^{(0)}(u)=0 \,,
\end{equation}
with the leading order transfer matrix given by
\begin{equation}
t^{(0)}_2(u)=2 \, u^2+q^{(0)}_2 \,, \qquad q^{(0)}_2=-M(M+1)-\tfrac{1}{2} \,.
\end{equation}
The solution to this equation is given in terms of the continuous Hahn polynomial
\begin{equation}\label{BaxterQ LO}
Q^{(0)}(u)={}_3 F_2\left(\left. \begin{array}{c}
-M, \ M+1,\ \frac{1}{2}+iu \\
1,\  1 \end{array}
\right| 1\right) \,.
\end{equation}
It is rather straightforward to compute the anomalous dimension using the above explicit formula for the Baxter function
\begin{equation}
\gamma^{(0)}=8\,S_1(M)\,.
\end{equation}
At next-to-leading order the Baxter equation can be obtained by expanding the corresponding
asymptotic all-loop equation, see \cite{BelitskyBaxter}, and is given by
\begin{eqnarray}\label{Baxter NLO}
\left(u+\tfrac{i}{2} \right)^2 Q^{(1)}(u+i)&+&\left(u-\tfrac{i}{2}\right)^2 Q^{(1)}(u-i)
 -t^{(0)}_2(u)Q^{(1)}(u)= \\
=\left(2- i \, \tfrac{\gamma^{(0)}}{2}(u+\tfrac{i}{2}) \right)Q^{(0)}(u+i)
 &+&\left(2+ i \, \tfrac{\gamma^{(0)}}{2}(u-\tfrac{i}{2}) \right)Q^{(0)}(u-i)
+t^{(1)}_2 (u)Q^{(0)}(u) \,, \nonumber
\end{eqnarray}
with $ t^{(1)}_2=-(\,4+\tfrac{1}{2}\gamma^{(0)}(2M+1))$.
In order to solve the above equation one is lead to consider the following class of continuous Hahn polynomials
\begin{equation} \label{ypsilon}
y(u)={}_3 F_2\left(\left. \begin{array}{c}
-n, \ n+a+b+c+d-1,\ a+iu \\
a+c,\  a+d \end{array}
\right| 1\right)\,,
\end{equation}
which satisfy the  difference equation \cite{Orthogonal Polynomials}
\begin{equation}\label{Difference Equation Hahn}
n(n+a+b+c+d-1) \, y(u)=B(u) \, y(u+i)-[B(u)+D(u)]y(u)+D(u) \, y(u-i)
\end{equation}
where
\begin{equation}
B(u)=(c-iu)(d-iu) \,, \qquad D(u)=(a+iu)(b+iu) \nonumber \,.
\end{equation}
By a deformation of the one-loop solution \eqref{BaxterQ LO} we understand a suitable 
choice of $a,b,c$ and $d$, such that when the deformation
parameters are set to zero, the function \eqref{ypsilon} coincides with \eqref{BaxterQ LO}. 
One of such deformations is given by
\begin{equation}\label{QA}
Q^A_\delta(u)={}_3 F_2\left(\left. \begin{array}{c}
-M, \ M+1+2\delta,\ \frac{1}{2}+iu \\
1+\delta,\  1 \end{array}
\right| 1\right) \,.
\end{equation}
Upon differentiating once w.r.t.~$\delta$ at $\delta=0$ the resulting difference equation reads
\begin{eqnarray}
&& \left(u+\tfrac{i}{2} \right)^2  Q^{A}_{0}{}'(u+i)+\left(u-\tfrac{i}{2}\right)^2 Q^{A}_{0}{}'(u-i)
 -t^{(0)}_2(u) Q^{A}_{0}{}'(u)= \nonumber\\
&&\quad =-i(u+\tfrac{i}{2}) \, Q^{(0)}(u+i)+i(u-\tfrac{i}{2})\, Q^{(0)}(u-i)
-(2M+1) \, Q^{(0)}(u) \,.
\end{eqnarray}
In the above formula we have used the following notation
\begin{equation}
\frac{\p}{\p\delta}\,\left. Q^{A}_{\delta}(u)\right|_{\delta=0} = Q^{A}_{0}{}'(u)
\quad {\mathrm{and}} \quad
\left. Q^{A}_{\delta}(u)\right|_{\delta=0} =Q^{(0)} \,.
\end{equation}
Thus using the linearity of \eqref{Baxter NLO} one can multiply \eqref{QA} with $\tfrac{\gamma^{(0)}}{2}$ to
obtain the part of $Q^{(1)}(u)$ which is proportional to the one-loop anomalous
dimension.

The missing parts of the full solution can be found by considering the following deformation of the one-loop solution
\begin{equation}\label{QB}
Q^B_\delta(u)={}_3 F_2\left(\left. \begin{array}{c}
-M, \ M+1,\ \frac{1}{2}+iu \\
1+\delta,\  1-\delta \end{array}
\right| 1\right) \,.
\end{equation}
The second derivative of the resulting difference equation evaluated at $\delta=0$
takes the form
\begin{eqnarray}
\left(u+\tfrac{i}{2} \right)^2 Q^{B}_{0}{}''(u+i)+\left(u-\tfrac{i}{2}\right)^2 Q^{B}_{0}{}''(u-i)
 -t^{(0)}_2(u) Q^{B}_{0}{}''(u)=2 \, Q^{(0)}(u)-2 \, Q^{(0)}(u+i)\,. \nonumber\\
\end{eqnarray}
To complete the solution one must note that a third type of the deformation must
be introduced such that the argument of the r.h.s. of
the above equation is shifted $u \to (u-i)$. From \eqref{Difference Equation Hahn} 
one infers that this is obtained
by interchanging $a \leftrightarrow c$
and  $b \leftrightarrow d$. Thus, the last term reads
\begin{equation}\label{QC}
Q^C_\delta(u)={}_3 F_2\left(\left. \begin{array}{c}
-M, \ M+1,\ \frac{1}{2}+iu+\delta \\
1+\delta,\  1+\delta \end{array}
\right| 1\right) \,.
\end{equation}
Since the one-loop solution \eqref{BaxterQ LO} is also a solution to the homogeneous part of the
two-loop Baxter equation \eqref{Baxter NLO} one can add it to $Q^{(1)}(u)$ with an arbitrary
coefficient function of $M$. To fix the function uniquely one notices that if the leading order
Baxter function is a polynomial of degree $M$, then the higher loop corrections are polynomials
of degree $M-1$ and for even distribution of roots, as in the case considered\footnote{This follows 
from expanding $u_k (g^2)$ in \eqref{Q expansion} in a power series in $g$.}, $M-2$.
Hence, one has to add the term $a(M)\, Q^{(0)}(u)$, where $-a(M)$ is
the ratio of the highest order term $u^M$ of $Q^{(1)}(u)$ and $Q^{(0)}(u)$ given by
\begin{equation}
a(M)=4\big(S_2(M)+4\,S_1(M)^2-2\,S_1(M)S_1(2M)\big)\,.
\end{equation}
Finally, the full next-to-leading order Baxter function with the appropriate
normalization\footnote{The global normalization of the full Baxter 
function $Q(u)=Q^{(0)}(u)+g^2\,Q^{(1)}(u)$ can be chosen arbitrarily.} is given by
\begin{eqnarray}\label{Q1 full}
Q^{(1)}(u)&=&a(M)\,{}_3 F_2\left(\left.
	\begin{array}{c}
		-M, \ M+1,\ \frac{1}{2}+iu \\
		1,\  1
	\end{array}
 	\right| 1\right) \nonumber\\
&&+ \, \frac{\gamma^{(0)}}{2} \, \frac{\p}{\p\delta} \, {}_3 F_2 \left. \left(\left.
	\begin{array}{c}
		-M, \ M+1+2\delta,\ \frac{1}{2}+iu \\
		1+\delta,\  1
	\end{array}
	\right| 1\right) \right|_{\delta=0} \nonumber\\
&&- \,  \phantom{\gamma^{(0)}} \, \frac{\p^2}{\p\delta^2} \, {}_3 F_2  \left. \left(\left.
	\begin{array}{c}
		-M, \ M+1,\ \frac{1}{2}+iu \\
		1+\delta,\  1-\delta
	\end{array}
	\right| 1 \right) \right|_{\delta=0}   \nonumber\\
&&- \,  \phantom{\gamma^{(0)}} \frac{\p^2}{\p\delta^2}\, {}_3 F_2  \left.\left(\left.
	\begin{array}{c}
		-M, \ M+1,\ \frac{1}{2}+iu+\delta \\
		1+\delta,\  1+\delta
	\end{array}
	\right| 1\right) \right|_{\delta=0}  \,.
\end{eqnarray}
With these results it is rather straightforward to compute the two-loop anomalous dimension in a closed
form, see \Appref{app:dimensions}. It is given
by the following known expression, where all harmonic sums are evaluated at argument $M$
\begin{equation}\label{eq:gamma1}
\gamma^{(1)}(M)=-16\,\big(S_3+S_{-3}-2\,S_{-2,1}+2\,S_1(S_2+S_{-2})\big)\,.
\end{equation}

\medskip

At three-loop order one finds the following perturbative Baxter equation for
$Q^{(2)}(u)$
\begin{eqnarray}\label{Baxter NNLO}
\left(u+\tfrac{i}{2} \right)^2 Q^{(2)}(u+i)&+&\left(u-\tfrac{i}{2}\right)^2 Q^{(2)}(u-i)
 -t^{(0)}_2(u)\,Q^{(2)}(u)= \nonumber\\
=\left(2- i\tfrac{\gamma^{(0)}}{2}(u+\tfrac{i}{2}) \right)Q^{(1)}(u+i)
 &+&\left(2+ i\tfrac{\gamma^{(0)}}{2}(u-\tfrac{i}{2}) \right)Q^{(1)}(u-i)
+t^{(1)}_2(u)Q^{(1)}(u)\nonumber\\
+P(u)\,Q^{(0)}(u+i) &+&P^*(u)\,Q^{(0)}(u-i)\nonumber\\
-\Big(\tfrac{\gamma^{(1)}}{2}(2M+1)&+&2K_2 + \left(u+\tfrac{i}{2}\right)^{-2}
                                   +\left(u-\tfrac{i}{2}\right)^{-2}\Big)Q^{(0)}(u) \,,
\end{eqnarray}
where $P(u)$ and $K_2=K_2(M)$ are given by
\begin{equation}
P(u)=\left(\frac{1}{\left(u+\frac{i}{2}\right)^2}
+\frac{i\, \gamma^{(0)}}{2\left(u+\frac{i}{2}\right)}
+  K_2 -\frac{i\gamma^{(1)}}{2} \left(u+\tfrac{i}{2} \right)\right), \, K_2(M)=\frac{\gamma^{(0)}(M)^2}{8}-\,4\,S_{-2}(M)\,.
\end{equation}

It is instructive to split the solution into five different sub-classes, which
are independent of each other due to the transcendental composition of the corresponding
inhomogeneities and can be analyzed separately as presented in Table \ref{table:class of solutions}.

\begin{table}[t]\centering
\begin{tabular}{||r|r||}
\hline
Index            &	Classification\\
\hline\hline
$\I\phantom{\I}$ &	polynomial parts of $Q_0$\\
$\II_a$ 	 &      terms of order $(\gamma^{(0)})^0$\\
$\II_b$  	 &	terms of order $(\gamma^{(0)})^1$\\
$\II_c$  	 &      terms of order $(\gamma^{(0)})^2$\\
$\III $  	 &      non-polynomial parts of $Q_0$\\
\hline
$\IV $  	 &      one-loop normalization\\
\hline
\end{tabular}
\caption{\sl{Classification of solutions}}
\label{table:class of solutions}
\end{table}

The classes $\I$, $\II_a$, $\II_b$ and $\II_c$ can be obtained by using the same arguments
as for the next-to-leading order analysis. Terms polynomial in $Q_0$ lead to
\begin{eqnarray}\label{Q2 I}
Q^{(2)}_{\I}(u)&=&\frac{\gamma^{(1)}+a(M)\,\gamma^{(0)}}{2}\,
 \frac{\p}{\p\delta} \, {}_3 F_2 \left. \left(\left.
	\begin{array}{c}
		-M, \ M+1+2\delta,\ \frac{1}{2}+iu \\
		1+\delta,\  1
	\end{array}
	\right| 1\right) \right|_{\delta=0} \nonumber\\
&&-\frac{K_2(M)+2\,a(M)}{2}\,
 \frac{\p^2}{\p\delta^2}  {}_3 F_2 \left .\left(\left.
	\begin{array}{c}
		-M, \ M+1,\ \frac{1}{2}+iu \\
		1+\delta,\  1 -\delta
	\end{array}
	\right| 1\right)\right|_{\delta=0} \nonumber\\
&&-\frac{K_2(M)+2\,a(M)}{2}\frac{\p^2}{\p\delta^2}
       \,{}_3 F_2 \left .\left(\left.
	\begin{array}{c}
		-M, \ M+1,\ \frac{1}{2}+iu+\delta \\
		1+\delta,\  1 +\delta
	\end{array}
	\right| 1\right)
\right|_{\delta=0}\,.
\end{eqnarray}
The $\II_a$ part of the solution is given by
\begin{eqnarray}\label{Q2 IIa}
Q^{(2)}_{\II_a}(u)&=&\frac{1}{6} \,\frac{\p^4}{\p\delta^4}
  \, {}_3 F_2 \left. \left(\left.
	\begin{array}{c}
		-M, \ M+1,\ \frac{1}{2}+iu \\
		1+\delta,\  1 -\delta
	\end{array}
	\right| 1\right) \right|_{\delta=0} \nonumber\\
&&+\frac{1}{6} \,  \frac{\p^4}{\p\delta^4}
   {}_3 F_2 \left .\left(\left.
	\begin{array}{c}
		-M, \ M+1,\ \frac{1}{2}+iu+\delta \\
		1+\delta,\  1 +\delta
	\end{array}
	\right| 1\right)\right|_{\delta=0} \nonumber\\
&&+\frac{\p^2}{\p\alpha^2}\,\frac{\p^2}{\p\beta^2}
       \,{}_3 F_2 \left .\left(\left.
	\begin{array}{c}
		-M, \ M+1,\ \frac{1}{2}+iu+\beta \\
		1+\alpha+\beta,\  1 -\alpha+\beta
	\end{array}
	\right| 1\right)
\right|_{\alpha,\beta=0}\,.
\end{eqnarray}
Polynomial inhomogeneities linear in $\gamma^{(0)}$ redound to the result
\begin{eqnarray}\label{Q2 IIb}
Q^{(2)}_{\II_b}(u)=-\frac{\gamma^{(0)}}{2} \Bigg(
  && \frac{\p}{\p\alpha}\,\frac{\p^2}{\p\beta^2}
  \, {}_3 F_2 \left. \left(\left.
	\begin{array}{c}
		-M, \ M+1+2\,\alpha,\ \frac{1}{2}+iu+\beta \\
		1+\alpha+\beta,\  1 +\beta
	\end{array}
	\right| 1\right) \right|_{\alpha,\beta=0} \nonumber\\
&&+\frac{1}{3} \,  \frac{\p^3}{\p\beta^3}
\, {}_3 F_2 \left. \left(\left.
	\begin{array}{c}
		-M, \ M+1,\ \frac{1}{2}+iu+\beta \\
		1+\beta,\  1 +\beta
	\end{array}
	\right| 1\right) \right|_{\beta=0} \nonumber\\
&&+ \frac{\p}{\p\alpha}\,\frac{\p^2}{\p\beta^2}
  \, {}_3 F_2 \left. \left(\left.
	\begin{array}{c}
		-M, \ M+1+2\,\alpha,\ \frac{1}{2}+iu \\
		1+\alpha+\beta,\  1 -\beta
	\end{array}
	\right| 1\right) \right|_{\alpha,\beta=0} \Bigg)\,.
%
%&&+\frac{1}{3} \,  \frac{\p^3}{\p\beta^3}
% \, {}_3 F_2 \left. \left(\left.
%	\begin{array}{c}
%		-M, \ M+1,\ \frac{1}{2}+iu \\
%		1+\beta,\  1 -\beta
%	\end{array}
%	\right| 1\right) \right|_{\beta=0} \Bigg)\,.
% THE LAST TERM CANCELS THE INHOMOGENEITY CORRECTLY, BUT IT IS ZERO AS IS THE INHOMOGENEITY!!
\end{eqnarray}
The term of $\Op((\gamma^{(0)})^2)$ is given by
\begin{equation}\label{Q2 IIc}
Q^{(2)}_{\II_c}(u)=\frac{\big(\gamma^{(0)}\big)^2}{8} \,\frac{\p^2}{\p\delta^2}
  \, {}_3 F_2 \left. \left(\left.
	\begin{array}{c}
		-M, \ M+1+2\delta,\ \frac{1}{2}+iu \\
		1+\delta,\  1
	\end{array}
	\right| 1\right) \right|_{\delta=0} \,.
\end{equation}
To find the solution for class $\III$, these are all the terms that contain the denominators
$\sfrac{1}{(u \pm i/2)}$, one needs to proceed more generally. The difference equation
satisfied by continuous Hahn polynomials does not allow to generate $\sfrac{1}{(u \pm i/2)}$
terms by the deformations previously introduced. However, all terms of type $\III$ add up to a polynomial and
one can make use of a general statement formulated in lemma $1$ to find the solution, see \Appref{app:Problems}
where also details of the calculation can be found. When brought together, the terms in class $\III$ can be
expressed by a sum of two polynomials. The first part is given by a hypergeometric polynomial, and the
corresponding solution can be easily found using the techniques discussed in this section
\begin{eqnarray} \label{QIII_a}
Q^{(2)}_{\III_a}(u)&=& \,  \frac{\gamma^{(0)}}{4}\, \frac{\p^2}{\p\alpha^2}\frac{\p}{\p\beta} \, {}_4 F_3 \left. \left(\left.
	\begin{array}{c}
		-M, \ M+1,\ \tfrac{1}{2}+iu ,\ 1\\
		1+\alpha,\  1-\alpha+\beta ,\ 1-\beta \end{array}
	\right| 1\right)\right|_{\alpha,\beta=0} \nonumber\\
&&+\,  \frac{\gamma^{(0)}}{4}\, \frac{\p^2}{\p\alpha^2}\frac{\p}{\p\beta} \, {}_4 F_3 \left. \left(\left.
	\begin{array}{c}
		-M, \ M+1,\ \tfrac{1}{2}-iu ,\ 1\\
		1+\alpha,\  1-\alpha+\beta ,\ 1-\beta \end{array}
	\right| 1\right)\right|_{\alpha,\beta=0}\,.
\end{eqnarray}
On the other hand, the second polynomial is much more complicated and to find the solution one has
to apply lemma $1$ from \Appref{app:Problems}
\begin{equation}\label{QIII_b}
Q^{(2)}_{\III_b}(u)=\sum_{k=0}^{M}\frac{m_k \, R(k,M)}{k!}\,
	          \big((\tfrac{1}{2} + iu)_k+(\tfrac{1}{2} - iu)_k\big)\,,
\quad m_k=\sum_{j=1}^{k}\frac{a_{j-1}\,(j-1)!}{j^2 \, R(j,M)}\,.
\end{equation}
See \appref{App:Mellin} and \appref{app:Problems} for the definitions of $R(j,M)$ and $a_j$.
Finally the solution to the homogeneous equation, i.e.~the leading order solution with the correct
normalization is fixed again by imposing the proper degree reduction of $Q_2(u)$ such that
$Q_2(u)$ is a polynomial of degree $M-2$. The normalization is computed in the same fashion
as in the two-loop case. Hence $Q^{(2)}_{\IV}$ is given by
\begin{equation}
Q^{(2)}_{\IV}(u)=a^{(2)}(M)\,{}_3 F_2\left(\left.
	\begin{array}{c}
		-M, \ M+1,\ \frac{1}{2}+iu \\
		1,\  1
	\end{array}
 	\right| 1\right),
\end{equation}
with
\begin{eqnarray}
a^{(2)}(M)&=&\frac{8}{3}\Big(52S_{1}(M)^4 - S_{-3}(M)\big( 9S_{1}(M)-6S_{1}(2M) \big) - 48S_{1}(M)^3 S_{1}(2M)
\nonumber\\
	 &&-3S_{-2}(M)\big(7S_{1}(M)^2-4S_{1}(M)S_{1}(2M)+S_{2}(M)\big)+3S_{1}(M)^2 \big(4S_{1}(2M)^2 \nonumber\\
	 &&-5S_{2}(M)+4S_{2}(2M) \big) - 6S_{4}(M) + 3\big(S_{2}(M)^2 +2S_{1}(2M)(S_{3}(M)-2S_{-2,1}(M))\big)\nonumber\\
	 &&-2S_{1}(M)\big(11S_{3}(M)-9S_{-2,1}(M) \big)\Big)-2m_M\,.
\end{eqnarray}
The anomalous dimension obtained from this result in terms of nested harmonic sums with
argument $M$ is given by
\begin{eqnarray} \label{twisttwothreeloops}
\gamma^{(2)}(M)=&&64\Big(2S_{-5}+2S_{5}-4S_{-4,1}-2S_{-3,-2}-S_{-3,2}-2S_{-2,-3}-8S_{1,-4}
-4S_{1,4}\nonumber\\
&&-9S_{2,-3}-5S_{2,3}-2S_{3,-2}-5S_{3,2}+2\big(-2S_{4,1}+S_{-2,-2,1}+S_{-2,1,-2}+4S_{1,-3,1}
\nonumber\\
&&+S_{1,-2,-2}+S_{1,-2,2}+6S_{1,1,-3}+2S_{1,1,3}+2S_{1,2,-2}+2S_{1,2,2}+2S_{1,3,1}+3S_{2,-2,1}\nonumber\\
&&+2(S_{2,1,-2}
+S_{2,1,2}+S_{2,2,1}+S_{3,1,1}-2S_{1,1,-2,1}) \big) \Big).
\end{eqnarray}
It agrees precisely with the conjecture of \cite{Kotikov:2004er}, upon changing the basis of the harmonic sums.
%%%%%%%%%%%%%%%%%%%%%%%%%%%%%%%%%%%%%%%%%%%%%%%%%%%%%%%%%%%%%%%%%%%%
\section{Twist-three operators}
A special class of maximal helicity twist-three operators in QCD has also
found to be integrable to the first loop correction \cite{Baryon integrability} and the relation of the solution
to Wilson polynomials has been established by the Baxter approach in \cite{Derkachov:1999ze}.

In this section we will apply the methods of the previous one to obtain the three-loop expression
for the Baxter function of twist-three operators of $\superN=4$ SYM
\begin{equation}\label{Twist 3 op}
\Tr \left( \fldD^{s_1} \fldZ\, \fldD^{s_2} \, \fldZ \fldD^{s_3} \, \fldZ\,\right) + \ldots\,, \quad
s_1+s_2+s_3=M\,.
\end{equation}
The leading order Baxter function is given by the Wilson polynomial \cite{Dressing&Wrapping}
\begin{equation}\label{BaxterQ LO T3}
Q^{(0)}(u)={}_4 F_3\left(\left. \begin{array}{c}
-\tfrac{M}{2}, \ \tfrac{M}{2}+1,\ \tfrac{1}{2}+iu ,\ \tfrac{1}{2}-iu\\
1,\  1 ,\ 1 \end{array}
\right| 1\right) \,,
\end{equation}
from which one computes the anomalous dimension
\begin{equation}
\gamma^{(0)}_3=8\,S_1(\tfrac{M}{2})\,.
\end{equation}

The Baxter equation for the twist-three operators \eqref{Twist 3 op} to two-loops is given by
\begin{eqnarray}
&\phantom{=}&\left(u+\tfrac{i}{2} \right)^3 Q^{(1)}(u+i)+\left(u-\tfrac{i}{2}\right)^3 Q^{(1)}(u-i)
-t^{(0)}_3(u)Q^{(1)}(u)=t^{(1)}_3 Q^{(0)}(u)\\
&&+\left( 3-\, i \, \tfrac{\gamma^{(0)}_3}{2} \left(u+\tfrac{i}{2} \right) \right)
\left(u+\tfrac{i}{2} \right) Q^{(0)}(u+i)
+\left( 3+\, i \, \tfrac{\gamma^{(0)}_3}{2} \left(u-\tfrac{i}{2} \right) \right)
\left(u-\tfrac{i}{2} \right) Q^{(0)}(u-i)\,,\nonumber
\end{eqnarray}
where
\begin{equation}
t^{(0)}_3(u)=2u^3-(M^2+2M+\tfrac{3}{2})u \,, \quad
t^{(1)}_3(u)=\big(-\gamma^{(0)}_3(M+1)-6\big)u \,.
\end{equation}
Using the same techniques as for the twist-two case one finds the final result to be given by
\begin{eqnarray}
Q^{(1)}(u)&=&c(M)
	{}_4 F_3\left(\left.
	\begin{array}{c}
		-\tfrac{M}{2}, \ \tfrac{M}{2}+1,\ \tfrac{1}{2}+iu ,\ \tfrac{1}{2}-iu\\
		1,\  1 ,\ 1 \end{array}
	\right| 1\right) \nonumber\\
&&+ \, \frac{\gamma^{(0)}_3}{4} \, \frac{\p}{\p\delta} \, {}_4 F_3 \left. \left(\left.
	\begin{array}{c}
		-\tfrac{M}{2}, \ \tfrac{M}{2}+1+2\delta,\ \tfrac{1}{2}+iu ,\ \tfrac{1}{2}-iu\\
		1+\delta,\  1+\delta ,\ 1 \end{array}
	\right| 1\right)\right|_{\delta=0} \nonumber\\
&&- \, \phantom{\gamma}\frac{3}{2} \, \frac{\p^2}{\p\delta^2} \, {}_4 F_3 \left. \left(\left.
	\begin{array}{c}
		-\tfrac{M}{2}, \ \tfrac{M}{2}+1,\ \tfrac{1}{2}+iu ,\ \tfrac{1}{2}-iu\\
		1+\delta,\  1-\delta ,\ 1 \end{array}
	\right| 1\right)\right|_{\delta=0}\,,
\end{eqnarray}
with the normalization function $c(M)$ given by
\begin{equation}\label{c(M)}
c(M)=\tfrac{\gamma^{(0)}_3}{4}\left( 4S_1(\tfrac{M}{2})-2S_1(M)\right)+3\,S_2(\tfrac{M}{2})\,.
\end{equation}
The anomalous dimension computed from these closed expressions,
see \appref{app:dimtwist3}, is given by
\begin{equation}\label{gamma1 twist3}
\gamma^{(1)}_3(M)=-8 \bigg(S_{3}(\tfrac{M}{2})+2S_{1}(\tfrac{M}{2})S_{2}(\tfrac{M}{2})\bigg)\,.
\end{equation}
as has been guessed in \cite{Dressing&Wrapping, Beccaria m=0}.

\medskip

The NNLO correction to the Baxter function can be found as the solution of
\begin{eqnarray}\label{Baxter NNLO twist3}
&&\left(u+\tfrac{i}{2} \right)^3 Q^{(2)}(u+i)+\left(u-\tfrac{i}{2}\right)^3 Q^{(2)}(u-i)
 -t_3^{(0)}(u)\,Q^{(2)}(u)= \\
&&=\left(3- i\tfrac{\gamma^{(0)}_3}{2}(u+\tfrac{i}{2}) \right)(u+\tfrac{i}{2})Q^{(1)}(u+i)
 +\left(3+ i\tfrac{\gamma^{(0)}_3}{2}(u-\tfrac{i}{2}) \right)(u-\tfrac{i}{2})Q^{(1)}(u-i)\nonumber\\
&&\phantom{=}+t_3^{(1)}(u)Q^{(1)}(u)+P_3(u)\,Q^{(0)}(u+i) +P_3^*(u)\,Q^{(0)}(u-i)-\big(\gamma^{(1)}_3(M+1)+2K_3 \big)Q^{(0)}(u)\nonumber\,,
\end{eqnarray}
where $P_3(u)$ and $K_3=K_3(M)$ are given by
\begin{equation}
P_3(u)=\left(i\, \gamma^{(0)}_3
+  K_3\left(u+\tfrac{i}{2} \right)\ -\frac{i\gamma^{(1)}_3}{2} \left(u+\tfrac{i}{2} \right)^2\right),
\,K_3(M)=\frac{\gamma^{(0)}_3(M)^2}{8}\,.
\end{equation}
Following the same method outlined in section \ref{Twist-two operators} and considering
each part of the solution separately according to table \ref{table:class of solutions} it is
straightforward to find the complete solution. The normalization function $c^{(2)}$ for the homogeneous
solution is again fixed by degree reduction i.e. $Q^{(2)}$ is a polynomial of degree $(M-2)$,
\begin{eqnarray}
c^{(2)}(M)&=&32S_1(\tfrac{M}{2})^4 - 32S_1(\tfrac{M}{2})^3 S_1(M) + \tfrac{9}{2}S_2(\tfrac{M}{2})^2
	    +8S_1(\tfrac{M}{2})^2\big(S_1(M)^2+S_2(M)\big) \nonumber\\
	 && +4S_1(M)S_3(\tfrac{M}{2}) - 4S_1(\tfrac{M}{2})\big(S_1(M) S_2(\tfrac{M}{2})+3S_3(\tfrac{M}{2}) \big)
           -\tfrac{9}{2}S_4(\tfrac{M}{2})\,.
\end{eqnarray}

The three-loop Baxter function is given by
\begin{eqnarray}
Q^{(2)}(u)&=&c^{(2)}(M)
	{}_4 F_3\left(\left.
	\begin{array}{c}
		-\tfrac{M}{2}, \ \tfrac{M}{2}+1,\ \tfrac{1}{2}+iu ,\ \tfrac{1}{2}-iu\\
		1,\  1 ,\ 1 \end{array}
	\right| 1\right) \nonumber\\
&&+ \, \frac{\gamma^{(1)}_3+c(M)\gamma^{(0)}_3}{4} \, \frac{\p}{\p\delta} \, {}_4 F_3 \left. \left(\left.
	\begin{array}{c}
		-\tfrac{M}{2}, \ \tfrac{M}{2}+1+2\delta,\ \tfrac{1}{2}+iu ,\ \tfrac{1}{2}-iu\\
		1+\delta,\  1+\delta ,\ 1 \end{array}
	\right| 1\right)\right|_{\delta=0} \nonumber\\
&&- \,  \frac{K_3(M)+3\,c(M)}{2}\, \frac{\p^2}{\p\delta^2} \, {}_4 F_3 \left. \left(\left.
	\begin{array}{c}
		-\tfrac{M}{2}, \ \tfrac{M}{2}+1,\ \tfrac{1}{2}+iu ,\ \tfrac{1}{2}-iu\\
		1+\delta,\  1-\delta ,\ 1 \end{array}
	\right| 1\right)\right|_{\delta=0}\nonumber\\
&&+ \,  \frac{3}{8}\, \frac{\p^4}{\p\delta^4} \, {}_4 F_3 \left. \left(\left.
	\begin{array}{c}
		-\tfrac{M}{2}, \ \tfrac{M}{2}+1,\ \tfrac{1}{2}+iu ,\ \tfrac{1}{2}-iu\\
		1+\delta,\  1-\delta ,\ 1 \end{array}
	\right| 1\right)\right|_{\delta=0}\nonumber\\
&&- \,  \frac{3}{8}\gamma^{(0)}_3\, \frac{\p}{\p\alpha}\frac{\p^2}{\p\beta^2} \, {}_4 F_3 \left. \left(\left.
	\begin{array}{c}
		-\tfrac{M}{2}, \ \tfrac{M}{2}+1+2\alpha,\ \tfrac{1}{2}+iu ,\ \tfrac{1}{2}-iu\\
		1+\alpha,\  1+\alpha+\beta ,\ 1-\beta \end{array}
	\right| 1\right)\right|_{\alpha,\beta=0}\nonumber\\
&&+ \,  \frac{1}{8}\gamma^{(0)}_3\, \frac{\p^2}{\p\alpha^2}\frac{\p}{\p\beta} \, {}_4 F_3 \left. \left(\left.
	\begin{array}{c}
		-\tfrac{M}{2}, \ \tfrac{M}{2}+1,\ \tfrac{1}{2}+iu ,\ \tfrac{1}{2}-iu\\
		1+\alpha,\  1-\alpha+\beta ,\ 1-\beta \end{array}
	\right| 1\right)\right|_{\alpha,\beta=0}\nonumber\\
&&+ \, \frac{(\gamma^{(0)}_3)^2}{32} \, \frac{\p^2}{\p\delta^2} \, {}_4 F_3 \left. \left(\left.
	\begin{array}{c}
		-\tfrac{M}{2}, \ \tfrac{M}{2}+1+2\delta,\ \tfrac{1}{2}+iu ,\ \tfrac{1}{2}-iu\\
		1+\delta,\  1+\delta ,\ 1 \end{array}
	\right| 1\right)\right|_{\delta=0} \nonumber\\
&&+ \, \frac{(\gamma^{(0)}_3)^2}{32} \, \frac{\p^2}{\p\delta^2} \, {}_4 F_3 \left. \left(\left.
	\begin{array}{c}
		-\tfrac{M}{2}, \ \tfrac{M}{2}+1,\ \tfrac{1}{2}+iu ,\ \tfrac{1}{2}-iu\\
		1+\delta,\  1-\delta ,\ 1 \end{array}
	\right| 1\right)\right|_{\delta=0}.
%
%&&+ \,  \frac{\gamma^{(0)}_3}{2}\, \frac{\p^2}{\p\alpha^2}\frac{\p}{\p\beta} \, {}_4 F_3 \left. \left(\left.
%	\begin{array}{c}
%		-\tfrac{M}{2}, \ \tfrac{M}{2}+1,\ \tfrac{1}{2}+iu ,\ \tfrac{1}{2}-iu\\
%		1+\alpha,\  1-\alpha+\beta ,\ 1-\beta \end{array}
%	\right| 1\right)\right|_{\alpha,\beta=0}\,.
\end{eqnarray}
For the three-loop anomalous dimension one obtains, again see \Appref{app:dimtwist3} for details,
\begin{equation}\label{gamma2 twist3}
\gamma^{(2)}(M)=8\big(S_{5}-2S_{1,4}-6S_{2,3}-10S_{3,2}-6S_{4,1}+8(S_{1,2,2}+S_{2,1,2}+S_{2,2,1}+S_{1,3,1}+S_{3,1,1}) \big),
\end{equation}
with all sums evaluated at $\sfrac{M}{2}$. The result coincides\footnote{Note,
that we choose to write the harmonic sums in the canonical basis.} with the
conjecture of \cite{Dressing&Wrapping, Beccaria m=0}.

It is important to note, that the three-loop Baxter equation for the twist-three operators, 
in contradistinction to the case of twist-two operators, \textit{does not} contain superficially 
non-polynomial parts and thus the solution can be found in terms of deformations of the one-loop 
solution only. Although the wrapping problem for twist-two operators starts from the four-loop order, 
this superficial breakdown of the polynomiality of the Baxter equation may signalize its 
incorrectness at the next order. The same applies for the twist-three operators, where the 
corresponding four-loop Baxter equation contains rational functions, though the solution should 
correctly reproduce the corresponding anomalous dimension \cite {Dressing&Wrapping}.
%%%%%%%%%%%%%%%%%%%%%%%%%%%%%%%%%%%%%%%%%%%%%%%%%%%%%%%%%%%%%%%%%%%%%%
\section{Summary and Outlook}\label{sec:Conclusion}

We have shown how to solve the two- and three-loop Baxter equation for a special
subset of operators. With the explicit form of the Baxter
function we were able to reproduce the known results based on
Feynman calculus \cite{Kotikov:2003fb} and to prove the three-loop
conjecture \cite{Kotikov:2004er} for twist two-operators that has e.g.~been
used to check the field theory solution for the three-loop planar dilatation generator
obtained by algebraic methods in \cite{Zwiebel:2008gr}.
Likewise, we gave a proof for the anomalous dimensions of twist-three operators
that were conjectured in \cite{Dressing&Wrapping, Beccaria m=0}.
Besides our approach to find analytic solutions to the Baxter equation there are
also techniques to obtain such solutions directly from Bethe ansatz equations, see
\cite{Nestin&Dressing} and especially \cite{Sasaki:2008ds} and references therein.

Due to the nature of the mechanism one can trace back all
different contributions to the anomalous dimension to the corresponding inhomogeneity
of the perturbative Baxter equation. As such it would be of great importance
to generalize the successfull application of TBA \cite{Bajnok:2008bm} of the
four-loop Konishi \cite{four-loop Konishi}, i.e.~twist-two $M=2$,
operator to all twist-two operators. The knowledge of the correct four-loop result
of twist-two operators in a finite volume could than be used to analyze and, if possible, to fix
the four-loop Baxter and hence Bethe equations. A detailed application of our methods to
the next order should also reveal a different structure appearing in the Baxter equation
that renders the twist-two solution incorrect but, in turn, should give the right result for twist-three.

%%%%%%%%%%%%%%%%%%%%%%%%%%%%%%%%%%%%%%%%%%%%%%%%%%%%%%%%%%%%%%%%%%%%%%%
\subsection*{Acknowledgments}
We would like to thank Matthias Staudacher for drawing our attention to this
problem. During the completion of this work we benefited from discussions with
Matteo Beccaria, Niklas Beisert, Lev Lipatov, Carlo
Meneghelli, Matthias Staudacher and Arkady Tseytlin. A.~R.~and S.~Z.~
like to thank Nathan Berkovits and IFT S\~ao Paulo for kind hospitality, where part of this project has been completed.
S.~Z.~thanks the Galileo Galilei Institute for
Theoretical Physics for the hospitality and the INFN for partial support
during the completion of this work. A.~V.~K.~was supported by RFBR grant 07-02-00902-a.
%%%%%%%%%%%%%%%%%%%%%%%%%%%%%%%%%%%%%%%%%%%%%%%%%%%%%%%%%%%%%%%%%%%%%%
\appendix
%%%%%%%%%%%%%%%%%%%%%%%%%%%%%%%%%%%%%%%%%%%%%%%%%%%%%%%%%%%%%%%%%%%%%%
\section{Solution in Mellin space}\label{App:Mellin}
In this appendix we will show how to derive the one and two-loop solution
of the Baxter equation for twist two operators in Mellin space following
\cite{Faddeev Korchemsky}. We consider the Mellin transform $Q^{(i)}(\omega)$
of the Baxter function $Q^{(i)}(u)$
\begin{equation}\label{app:Mellin transform}
Q^{(i)} (u) ~=~ \int^{\infty}_{0} d \omega \, \omega^{iu-1} Q^{(i)} (\omega)\,.
\end{equation}
In what follows we will make use of the relations
\begin{eqnarray}\label{app:Mellin ingredients}
Q^{(i)} (u+ai) &=&
\int^{\infty}_{0} d \omega \, \omega^{iu-1} \,
\Bigl\{ \omega^{-a} Q^{(i)} (\omega) \Bigr\}, \nonumber \\
(u+bi)^J Q^{(i)} (u+ai) &=& \int^{\infty}_{0} d \omega \, \omega^{iu-1} \, \omega^{-b}
\Bigl(i\omega \frac{d}{d\omega}\Bigr)^J \Bigl\{ \omega^{b-a} Q^{(i)} (\omega) \Bigr\}\,.
\end{eqnarray}
{\bf One-loop.} The differential equation we obtain from the leading order Baxter equation
\eqref{Baxter LO} reads
\begin{eqnarray}\label{app:LO Baxter Mellin}
\left\{
\frac{(\omega-1)^2}{\omega} \Bigl(i\omega \frac{d}{d\omega}\Bigr)^2 + i\, \frac{\omega^2-1}{\omega} \Bigl(i\omega \frac{d}{d\omega}\Bigr)
- q_{2}^{(0)} - \frac{\omega^2+1}{4 \, \omega} \right\} Q^{(0)} (\omega) =0\,.
\end{eqnarray}
By means of the variable transformation $z=1/(1-\omega)$ and by applying
the Faddeev-Korchemsky substitution, see \cite{Faddeev Korchemsky},
$Q^{(i)} (\omega) \to Q^{(i)} (-\omega)$ in \eqref{app:Mellin transform}
we replace \eqref{app:Mellin transform} and \eqref{app:LO Baxter Mellin} by
\begin{equation}\label{app:Mellin transform z}
Q^{(i)} (u) = \frac{-i\pi}{\Gamma(iu)\Gamma(1-iu)} \, \int^{1}_{0} dz \, (1-z)^{iu-1} z^{-iu-1} \, Q^{(i)} (z)\,,
\end{equation}
\begin{equation}\label{app: LO Baxter Mellin z}
\left\{ z(1-z) \Bigl(\frac{d}{dz}\Bigr)^2
- q_{2}^{(0)} + \frac{1-2z(1-z)}{4z(1-z)} \right\} Q^{(0)} (z) = 0\,,
\end{equation}
respectively. It is convenient to introduce new functions $\bar{Q}^{(i)}(z)$
\begin{equation}\label{def:Q bar}
Q^{(i)}(z) = \sqrt{z(1-z)} \, \bar{Q}^{(i)}(z),
\end{equation}
for which \eqref{app: LO Baxter Mellin z} takes a simpler form
\begin{equation}\label{app:LO Baxter Mellin z bar}
\left\{ z(1-z) \Bigl(\frac{d}{dz}\Bigr)^2 + (1-2z) \frac{d}{dz}
- q_{2}^{(0)} - \frac{1}{2} \right\} \bar{Q}^{(0)} (z) =0,
\end{equation}
where we have omitted the factor $\sqrt{z(1-z)}$. It should be noted, that 
both $\bar{Q}^{(0)}(z)$ and $\bar{Q}^{(0)}(1-z)$ satisfy the above equation. 
Being a differential equation of the second order, equation \eqref{app:LO Baxter Mellin z bar} 
has two algebraically independent solutions and both, $\bar{Q}^{(0)}(z)$  and $\bar{Q}^{(0)}(1-z)$, 
can be written as linear combination of these two basis vectors. In particular, 
one may impose symmetry properties on the solutions. We choose the two independent 
solutions to obey\footnote{In what follows we assume even values of $M$.} $Q^{(0)}(1-z)=Q^{(0)}(z)$ and $Q^{(0)}_B(1-z)=-Q^{(0)}_B(z)$. The first solution can be 
found with the ansatz
\begin{equation}\label{app:LO Sol}
\bar{Q}^{(0)}(z) = \sum^{\infty}_{k=0} \, C_k \, z^{k+\alpha}
\end{equation}
with arbitrary coefficients $C_k$ and $\alpha$.
The beginning of the resulting recurrence leads to the consistency condition $\alpha=0$
and comparing terms with same powers of $z$, we find the relation
\begin{equation}\label{app:c_k rec}
(k+1)^2C_{k+1}=(k-M)(k+M+1)C_k\,.
\end{equation}
This first order recurrence is solved by
\begin{equation}
C_k= R(k,M)\,C_0 \,,
\end{equation}
where
\begin{equation}\label{app:R(k,M)}
R(k,M)= \frac{\Gamma(k+M+1)\Gamma(k-M)}{(k!)^2\,\Gamma(M+1)\Gamma(-M)} =
\frac{(-1)^k\,\Gamma(k+M+1)}{(k!)^2\,\Gamma(M+1-k)}\,.
\end{equation}
Thus, one solution to the Baxter equation \eqref{app:LO Baxter Mellin z bar}
reads
\begin{equation}\label{app:Sol z bar}
\bar{Q}^{(0)}(z) = C_0 \,\, {}_2 F_1\left(\left.
				\begin{array}{c}
				-M, \ M+1 \\ 1
				\end{array}
 				\right| z\right)\,.
\end{equation}
Using \eqref{app:Mellin transform z} we obtain\footnote{For the integration 
we make use of the invariance $\bar{Q}^{(0)}(z)=\bar{Q}^{(0)}(1-z)$}
the following final result in the variable $u$
\begin{equation}\label{app: LO Sol}
Q^{(0)}(u) = C_0  \,
\frac{i \pi \,\Gamma(\frac{1}{2}+iu)\Gamma(\frac{1}{2}-iu)}{\Gamma(iu)\Gamma(1-iu)}
\, {}_3 F_2\left(\left.
	\begin{array}{c}
	-M, \ M+1,\ \frac{1}{2}+iu \\
	1,\  1
	\end{array}
	\right| 1\right)\,.
\end{equation}
Note, that the factor containing gamma functions is just a phase and hence negligible.
%It is possible to find a second solution to the Baxter equation by a similar ansatz
%like \eqref{app:LO Sol} multiplied by logarithmic terms in $z$, however the
%resulting solution in $u$ space is not independent from the first. Despite this fact, it will be necessary
%to also obtain its representation.

The second solution can be found by including logarithmic terms in addition to \eqref{app:LO Sol}. Imposing the antisymmetry, we are looking for a solution of the following form
\begin{equation}\label{app:LO 2.Sol}
\bar{Q}^{(0)}_B(z) = \sum^{\infty}_{k=0}
		   \left(a_k \log z - a_k \log (1-z) +b_k\right) \, R(k,M) \, z^{k}\,.
\end{equation}
Terms of power $\log(z)$ and $-\log(1-z)$ lead to the relation
\begin{equation}
(k+1)^2  \, R(k+1,M)\, a_{k+1}=(k-M)(k+M+1)  \, R(k,M)\, a_k\,,
\end{equation}
which by definition of $R(k,M)$ in \eqref{app:R(k,M)} simply leads to
\begin{equation}
a_{k+1}=a_k \equiv a_0\,.
\end{equation}
All terms of $z^k$ lead to the intertwining relation for the
coefficients $a_k$ and $b_k$
\begin{equation}
2(k+1)\,R(k+1,M)\,a_{k+1}+(k+1)^2\,R(k+1,M)\,b_{k+1}=
	(k-M)(k+M+1) \,R(k,M)\,b_k\,,
\end{equation}
which can be solved straightforwardly
\begin{equation}
b_k=b_0 - 2 \, a_0 \,S_1(k)\,.
\end{equation}
Please note, that one can make use of the antisymmetry of the solution to fix $b_0$ to be
\begin{equation}
b_0=b_0 (M) =2 S_1 (M) a_0 \,.
\end{equation}
Upon integration of \eqref{app:LO 2.Sol} by means of taking a derivative of
the Beta integral
\begin{equation}\label{beta integral}
\int_0^1 dz \, z^{i u-1}(1-z)^{-i u-1}\sqrt{z(1-z)}z^k
 =\frac{1}{k!}\,(\tfrac{1}{2}+iu)_k\,
\Gamma(\tfrac{1}{2}+iu)\,\Gamma(\tfrac{1}{2}-iu)\,,
\end{equation}
the second solution is given by
\begin{equation}
Q^{(0)}_B(u)=\sum_{k=0}^{M}\frac{(-M)_k(1+M)_k(\frac{1}{2}+iu)_k}{(1)_k (1)_k\, k!}
\left(\Psi_0(\tfrac{1}{2}+iu+k)-\Psi_0(\tfrac{1}{2}-iu)-2S_1(k) +2  S_1(M)\right).
\end{equation}
The relation of this solution to the one already obtained can
be seen using the following identity for the Polygamma function
\begin{equation}\label{id:Psi_0}
\Psi_0(\tfrac{1}{2}+iu)- \Psi_0(\tfrac{1}{2}-iu)=i\pi \tanh(\pi u).
\end{equation}
It can then be written in the following form
\begin{equation}\label{app:2.Sol}
Q^{(0)}_B(u) = \frac{d}{d\delta}\left.
\, {}_3 F_2\left(\left.
	\begin{array}{c}
	-M, \ M+1,\ \frac{1}{2}+iu+\delta \\
	1+\delta,\  1+\delta
	\end{array}
	\right| 1\right)\right|_{\delta=0}\,+\big(2S_1(M) +i\pi \tanh(\pi u)\big)\,Q^{(0)}(u).
\end{equation}
The first part of this solution agrees with \eqref{app: LO Sol} when multiplied by the normalization factor 
$-2S_1(M)$ while the second part is nothing but \eqref{app: LO Sol} multiplied by a periodic function and a coefficient 
function of $M$.

Note, that there is still another representation for $Q^{(0)}$. Making a different choice of
variables $p=4z(1-z)$ to rewrite \eqref{app: LO Baxter Mellin z} and performing the same steps
of the computation one finds
\begin{equation}\label{app: LO Sol s=1/2}
Q^{(0)}(u) = C_0  \,
\frac{i \pi \,\Gamma(\frac{1}{2}+iu)\Gamma(\frac{1}{2}-iu)}{\Gamma(iu)\Gamma(1-iu)}
\, {}_4 F_3\left(\left.
	\begin{array}{c}
	-\frac{M}{2}, \ \frac{M+1}{2},\ \frac{1}{2}-iu,\ \frac{1}{2}+iu \\
	1,\ 1,\ \frac{1}{2}
	\end{array}
	\right| 1\right)\,.
\end{equation}
This representation is equal to \eqref{app: LO Sol}, see for example \cite{Prudnikov}.

\medskip

\noindent
{\bf Two-loops.} The differential equation that corresponds to the two-loop
Baxter equation \eqref{Baxter NLO} for the same choice of variables \eqref{def:Q bar}
reads
\begin{eqnarray}\label{app: NLO Baxter Mellin z}
&&\left\{ z(1-z) \Bigl(\frac{d}{dz}\Bigr)^2 + (1-2z) \frac{d}{dz}
- q_{2}^{(0)} - \frac{1}{2} \right\} \bar{Q}^{(1)} (z)  \nonumber\\
&& \phantom{\Big\{}= - \left\{ \frac{\gamma^{(0)}}{2}\left((1-2z)\frac{d}{dz} + 2M\right)
 + \frac{2}{z(1-z)}\right\}\bar{Q}^{(0)} (z) \,.
\end{eqnarray}
As the acting differential operator is linear we will analyze the two
inhomogeneous terms separately, starting with the part proportional to $\gamma^{(0)}$.
It can be rewritten as
\begin{equation}
-\left\{\frac{\gamma^{(0)}}{2}\left((1-2z)\frac{d}{dz}+2M\right)\right\}\bar{Q}^{(0)}(z)
=\frac{\gamma^{(0)}}{2} \sum_{k=0}^{M}B_k\, z^k \,,
\end{equation}
where the coefficients $B_k$ are given by
\begin{equation}
B_k=\frac{(k-M)(1+k-M)}{k+1} \, R(k,M) \, C_0\,.
\end{equation}
To find the first part of the solution $\bar{Q}^{(1)}_{A}(z)$ we make
the ansatz
\begin{equation}
\bar{Q}^{(1)}_{A}(z)=\sum_{k=0}^{M}D_k \, R(k,M) \, z^k \,,
\end{equation}
which leads to the condition
\begin{equation}
D_{k+1}=D_k+\frac{\gamma^{(0)}}{2} \, \frac{1+k-M}{(1+k)(1+k+M)} \, C_0 \,.
\end{equation}
The solution is given by
\begin{equation}
 D_{k}=D_0+\frac{\gamma^{(0)}}{2} \,\big(2\,S_1(k+M)-2\,S_1(M)-S_1(k) \big)\, C_0 \,.
\end{equation}
Thus, the first part of the result can be written in a compact form, noting that
\begin{equation}\label{app:Identity 1}
\frac{d}{d\delta}\left.\frac{(1+M+2\delta)_k}{(1+\delta)_k}\right|_{\delta=0}=
\frac{(1+M)_k}{(1)_k}\big(2\,S_1(k+M)-2\,S_1(M)-S_1(k) \big)\,,
\end{equation}
it is given by
\begin{eqnarray}
\bar{Q}^{(1)}_A(z) = D_0 \,\, {}_2 F_1\left(\left.
				\begin{array}{c}
				-M, \ M+1 \\ 1
				\end{array}
 				\right| z\right)
+  C_0 \,\frac{\gamma^{(0)}}{2}\,\frac{d}{d\delta}\left. {}_2 F_1\left(\left.
				\begin{array}{c}
				-M, \ M+1+2\,\delta \\ 1 +\delta
				\end{array}
 				\right| z\right)\right|_{\delta=0}\,.
\nonumber\\
\end{eqnarray}
In $u$-space the result is obtained according to \eqref{app:Mellin transform z}
and hence reads
\begin{eqnarray}\label{app:Q1}
Q^{(1)}_A(u)&=&\Lambda\,D_0 \, \,{}_3 F_2\left(\left.
	\begin{array}{c}
		-M, \ M+1,\ \frac{1}{2}+iu \\
		1,\  1
	\end{array}
 	\right| 1\right) \nonumber\\
&&+ \,\Lambda\, C_0 \, \frac{\gamma^{(0)}}{2} \, \frac{d}{d\delta} \, {}_3 F_2 \left. \left(\left.
	\begin{array}{c}
		-M, \ M+1+2\delta,\ \frac{1}{2}+iu \\
		1+\delta,\  1
	\end{array}
	\right| 1\right) \right|_{\delta=0} \,,
\end{eqnarray}
with, for completeness, the phase $\Lambda$ given by
\begin{equation}
\Lambda=\frac{i \pi \,\Gamma(\frac{1}{2}+iu)\Gamma(\frac{1}{2}-iu)}{\Gamma(iu)\Gamma(1-iu)}\,.
\end{equation}

Finally, let us focus on the second term to complete the
solution, i.e.
\begin{equation}\label{app: NLO Baxter Mellin 1/z term}
\left\{ z(1-z) \Bigl(\frac{d}{dz}\Bigr)^2 + (1-2z) \frac{d}{dz}
+ M(M+1) \right\} \bar{Q}^{(1)} (z)=
-\frac{2\,C_0}{z(1-z)}\sum_{k=0}^{M} \, R(k,M) \, z^k \,\,.
\end{equation}
One realizes that for this specific differential operator, we have to
make the ansatz that $\bar{Q}^{(1)}_{B_1}(z) \sim \log^2 z \sum_{k=0}^{M}a_k\,z^k$
to obtain any expression that contains $1/z$ terms. However, the resulting
expression also includes terms of order $\log^2 z$, $\log z$, $z^k$ such that
we need three consistency conditions to meet the requirement of the r.h.s.~of
\eqref{app: NLO Baxter Mellin 1/z term}. Therefore we should consider an ansatz
of the following form
\begin{equation}\label{app:NLO & 2.Sol}
\bar{Q}^{(1)}_{B}(z) = \sum^{\infty}_{k=0}
		   \left(a_k \left(\log z - \log (1-z)\right)^2
                       +b_k \left(\log z - \log (1-z)\right)
                       +c_k\right) \, R(k,M) \, z^{k}\,.
\end{equation}
Comparing the powers of $\log^2 z$, $\log z$ and $z^k$ respectively leads to the following
recurrences
\begin{eqnarray}\label{app:rec twist two last term}
a_{k+1} = a_k &\equiv& a_0 \,,\nonumber\\
b_{k+1} = b_k &-& \frac{4}{k+1}\,a_{0} \,,\nonumber\\
c_{k+1} = c_k &-& \frac{2}{k+1} \,b_{k+1}\, \quad
\mathrm{and} \quad a_0=-C_0\,.
\end{eqnarray}
The solutions to \eqref{app:rec twist two last term} are given by
\begin{equation}\label{app:recSol last term}
b_k=b_0 - 4 \, a_0 \, S_1(k), \quad
c_k=c_0-2 \, b_0 \,S_1(k)+ 8 \,a_0 \,S_{1,1}(k)\,.
\end{equation}
Note, that \eqref{app:NLO & 2.Sol} is the natural transcendental generalization
of the second solution of the leading order Baxter function \eqref{app:LO 2.Sol}
and as such its $b_0(1-2S_1(k))$ part leads to \eqref{app:2.Sol} and is not of
importance since the one-loop part will be fixed in the end by requirement of degree
reduction. For the same reason it is also dispensable to fix $c_0$ by symmetry requirements.
In addition writting the double index sum as
$2\,S_{1,1}(k)=S_1(k)^2+S_2(k)$ the result is given by
\begin{eqnarray}\label{app:Sol last term?}
Q^{(1)}_{B}(u)&=&\sum_{k=0}^{M}\frac{(-M)_k(1+M)_k(\frac{1}{2}+iu)_k}{(1)_k (1)_k\, k!}
\Big(\Psi_1(\tfrac{1}{2}+iu+k)+\Psi_1(\tfrac{1}{2}-iu)+2\, S_2(k)  \nonumber\\
&&+\big(\Psi_0(\tfrac{1}{2}+iu+k)-\Psi_0(\tfrac{1}{2}-iu)-2\, S_1(k)\big)^2+2\,S_2(k)\Big).
\end{eqnarray}
One of the terms containing $2\,S_2(k)$ can be written in a compact form by absorbing
it into a deformation of the Pochhammer symbol
\begin{equation}\label{app:Identity 2}
\frac{d^2}{d\delta^2}\left.\frac{1}{(1+\delta)_k (1-\delta)_k}\right|_{\delta=0}=
\frac{2}{(1)_k (1)_k} \, S_2(k)\,.
\end{equation}
Likewise as in the case of the second solution at leading order
we use the identity \eqref{id:Psi_0} and
\begin{equation}\label{id:Psi_1}
\Psi_1(\tfrac{1}{2}-iu)+ \Psi_1(\tfrac{1}{2}+iu)=\frac{\pi^2}{\cosh^2(\pi u)}\,,
\end{equation}
to obtain the following form
\begin{eqnarray}\label{app:Q1B}
 Q^{(1)}_{B}(u)&=&
%\pi^2\,Q^{(0)}(u)
%+2\,i\,\pi\tanh(\pi u)\,Q^{(0)}_B(u)
%\nonumber\\
-  C_0 \, \frac{d^2}{d\delta^2} \, {}_3 F_2 \left. \left(\left.
	\begin{array}{c}
		-M, \ M+1,\ \frac{1}{2}+iu+\delta \\
		1+\delta,\  1+\delta
	\end{array}
	\right| 1\right) \right|_{\delta=0}  \nonumber\\
&&- C_0 \, \frac{d^2}{d\delta^2} \, {}_3 F_2 \left. \left(\left.
	\begin{array}{c}
		-M, \ M+1,\ \frac{1}{2}+iu \\
		1+\delta,\  1-\delta
	\end{array}
	\right| 1\right) \right|_{\delta=0} \,.
\end{eqnarray}
We have neglected all one-loop parts and phases and fix the overall leading order
influence to the two-loop Baxter function by the fact that its degree
should be $M-2$. The final result
is given in \eqref{Q1 full}.
%%%%%%%%%%%%%%%%%%%%%%%%%%%%%%%%%%%%%%%%%%%%%%%%%%%%%%%%%%%%%%%%%%%%%%
\section{Anomalous dimensions}\label{app:dimensions}
To compute the anomalous dimension from the closed expressions
obtained, it is important to note that the twist-two and twist-three Baxter functions
are real and hence invariant under the map $u \to -u$. Furthermore the leading-order
$Q$-functions evaluated at $\pm \sfrac{i}{2}$ are equal to one,
i.e.~$Q^{(0)}(\pm \sfrac{i}{2})=1$.
%%%%%%%%%%%%%%%%%%%%%%%%%%%%%%%%%%%%%%%%%%%%%
\subsection{Twist-two operators}
{\bf Two-loops.} From \eqref{eq:gamma} one infers the following form of $\gamma^{(1)}$
\begin{equation}
\gamma^{(1)}(M)=A(M)+B(M)\,,
\end{equation}
where
\begin{eqnarray}
A(M)&=&i \Big(Q^{(0)}{}'''(u)+ 2\,Q^{(0)}{}'(u)^3
     -3\,Q^{(0)}{}'(u)\,Q^{(0)}{}''(u)\Big)\Big|_{u=-\frac{i}{2}}^{u=+\frac{i}{2}}
=A_{+} - A_{-}\,,\label{A}\\
B(M)&=&2\,i\Big(Q^{(1)}{}'(u)-Q^{(1)}(u)\,Q^{(0)}{}'(u) \Big)\Big|_{u=-\frac{i}{2}}^{u=+\frac{i}{2}}
=B_{+} - B_{-}\,.\label{B}
\end{eqnarray}
It is convenient to express $Q^{(0)}(u)$ in the following way
\begin{equation}\label{Q0 in binomials}
Q^{(0)}(u)=\sum_{k=0}^{M}\frac{b_{k,M}}{k!}(\tfrac{1}{2}+iu)_k\,,
\quad \mathrm{with} \quad b_{k,M}=(-1)^k \binom{M}{k}\binom{M+k}{k}\,.
\end{equation}
Taking the derivative and using some identities for nested harmonic numbers,
see \cite{Vermaseren:1998uu}, one finds the closed expression\footnote{Please, be reminded
that all states of $\alg{sl}(2)$ have even $M$.} for the
terms entering $A_-(M)$
\begin{eqnarray}
Q^{(0)}{}'(-\tfrac{i}{2})&=&i\sum_{k=1}^{M} b_{k,M} \,S_1(k)
			   =2\,i\,S_1(M)\,,\\
Q^{(0)}{}''(-\tfrac{i}{2})&=&i^2\sum_{k=1}^{M} b_{k,M}\, (2S_{1,1}(k)-2S_2(k))
			    =4\,i^2\big(2S_{1,1}-S_{2}+S_{-2}\big)\,,\\
Q^{(0)}{}'''(-\tfrac{i}{2})
&=&i^3\sum_{k=1}^{M} b_{k,M}\,(6S_{1,1,1}(k)-6S_{1,2}(k)-6S_{2,1}(k)+6S_3(k))\nonumber\\
%&=&i^3\sum_{k=0}^{M} b_{k,M}\,(S_1(k)^3-3\,S_1(k)\,S_2(k)+2\,S_3(k))\nonumber\\
&=&24\,i^3\big(2\,S_{1,1,1}-S_{1,2}-S_{2,1}+S_{1,-2}-S_{-2,1}\big)\,.
\end{eqnarray}
Hence, $A_-$ is given by
\begin{equation}
A_-(M)=8 \big(3\,S_{-3}(M)-S_{3}(M)-6\,S_{-2,1}(M) \big)\,.
\end{equation}
One of the sums that needs to be evaluated for $B_-$ is
\begin{equation}
Q^{(1)}(-\tfrac{i}{2})=a(M)+4\,S_{-2}(M)-8\,S_{1,1}(M)+4\,S_2(M)\,.
\end{equation}
To find the second part we are in need of the following auxiliary
formula
\begin{equation}
\sum_{k=1}^{M} b_{k,M}\,S_1(k+M)\,S_1(k)=
(-1)^M (8\, S_{1,1}(M)-5\,S_{2}(M))\,.
\end{equation}
After some algebra $Q^{(1)}{}'(-\tfrac{i}{2})$ is given by, harmonic sums evaluated
with argument $M$,
\begin{eqnarray}
Q^{(1)}{}'(-\tfrac{i}{2})&=&2\,i\,\big(\,a(M)\,S_1 - 4\,\big(2\,S_3+S_{-2,1}-S_{1,-2}+
S_1(\,S_{-2}+S_2)\nonumber\\
&& -3 \,(\,S_{1,2}+S_{2,1}-2\,S_{1,1,1})\big)\Big) \,,
\end{eqnarray}
such that $B_-$ reads
\begin{equation}
B_{-}(M)=16\big(\,S_3-S_{-3}+S_1(\,S_2+S_{-2})+2\,S_{-2,1}\big)\,.
\end{equation}
Due to the symmetry of the Baxter function, $A_+=-A_-$ and $B_+=-B_-$, the
two-loop anomalous dimension of twist-two operators is given by \eqref{eq:gamma1}.

\medskip

\noindent
{\bf Three-loops.} It is straightforward to expand the formula for $\gamma^{(2)}$
in terms of the perturbative Baxter functions. In order to find its close form we
need to evaluate the following sums of the leading-order result
\begin{eqnarray}
Q^{(0)}{}^{(4)}(-\tfrac{i}{2})
&=&i^4\sum_{k=1}^{M} b_{k,M}\,\left(24S_{1,1,1,1}(k)-12S_1(k)^2S_2(k)-12S_4(k)\right)\nonumber \\
&=&96\,i^4\big(2 ( 2S_{1,1,1,1}-S_{1,1,2}+S_{1,1,-2}
 -S_{1,2,1}-S_{1,-2,1}-S_{2,1,1}+S_{-2,1,1} ) \nonumber\\
&&+S_{2,2}-S_{2,-2}-S_{-2,2}+S_{-2,-2} \big) \,,
\end{eqnarray}
\begin{eqnarray}
Q^{(0)}{}^{(5)}(-\tfrac{i}{2})&=&960\,i^5 \big( 2 (2S_{1,1,1,1,1}-S_{2,1,1,1}-S_{1,2,1,1}-S_{1,1,2,1}-S_{1,1,1,2}+S_{1,1,1,-2}-S_{1,1,-2,1}\nonumber\\
&&+S_{1,-2,1,1}-S_{-2,1,1,1})+S_{1,2,2}+S_{2,1,2}+S_{2,2,1}-S_{1,-2,2}-S_{1,2,-2}\nonumber\\
&&+S_{1,-2,-2}+S_{-2,1,2}-S_{2,1,-2}-S_{-2,1,-2}+S_{-2,2,1}+S_{2,-2,1}+S_{-2,-2,1}\big)\,.
\end{eqnarray}
The NLO result contributes with the following terms
\begin{eqnarray}
Q^{(1)}{}''(-\tfrac{i}{2})&=&4\,i^2
\Big(a(M)(2S_{1,1}-S_{2}+S_{-2})-8\big(4(3S_{1,1,1,1}-S_{2,1,1}-S_{1,2,1}-S_{1,1,2})\nonumber\\
&&+2(S_{1,1,-2}+S_{1,-2,1}-S_{-2,1,1})+S_{1,3}+S_{3,1}-S_{1,-3}+S_{-3,1}+S_{2,2}\nonumber\\
&&-S_{2,-2}+S_{-2,2}-S_{-2,-2}\big) \Big) \\
Q^{(1)}{}'''(-\tfrac{i}{2})&=&24\,i^3
\Big( a(M)(2S_{1,1,1}-S_{1,2}-S_{2,1}+S_{1,-2}-S_{-2,1})-8 \big(20S_{1,1,1,1,1}\nonumber\\
&&-7 ( S_{1,1,1,2}+S_{1,1,2,1}+S_{1,2,1,1}+S_{2,1,1,1} ) +5S_{1,1,1,-2}+S_{1,1,-2,1}-3S_{1,-2,1,1}\nonumber\\
&&+S_{-2,1,1,1}+2S_{1,2,2}+2S_{2,1,2}+2S_{2,2,1}-2S_{1,2,-2}+S_{1,-2,2}-S_{1,-2,-2}\nonumber\\
&&-2S_{2,1,-2}-S_{-2,1,2}+S_{-2,1,-2}-S_{-2,2,1}-S_{-2,-2,1}+S_{1,1,3}+S_{1,3,1}+S_{3,1,1}\nonumber\\
&&-S_{1,1,-3}+S_{1,-3,1}-S_{-3,1,1}+S_{3,-2}+S_{-3,2}\big)\Big)\,.
\end{eqnarray}
$Q^{(2)}$ results in the contributions
\begin{eqnarray}
Q^{(2)}(-\tfrac{i}{2})&=& a^{(2)}(M)+4\big( 6S_{-4}+a(M)( S_{-2}+S_{2}-2S_{1,1} ) \big)
+2\big(4S_{4}-7S_{-3,1}-3S_{-2,2}\nonumber\\
&&-5S_{1,-3}-9S_{1,3}-S_{2,-2}-10S_{2,2}-9S_{3,1}+2\big(3S_{-2,1,1}+S_{1,1,-2}\nonumber\\
&&+6(S_{1,1,2}+S_{1,2,1}+S_{2,1,1}-4S_{1,1,1,1} )\big) \big) \nonumber\\
Q^{(2)}{}'(-\tfrac{i}{2})&=& 2i
\Big(8S_{-5}+a^{(2)}(M)S_{1}+4a(M)\big(S_{-3}-S_{3}+2(S_{2,1}+S_{1,2}-S_{-2,1}-3S_{1,1,1} )\big)
\nonumber\\
&&-8\big(4S_{5}+4S_{-4,1}+2S_{-3,-2}+7S_{-3,2}-S_{-2,3}-13S_{1,4}-7S_{2,-3}-20S_{2,3}\nonumber\\
&&-S_{3,-2}-20S_{3,2}-15S_{4,1}-4S_{-3,1,1}+3S_{-2,1,2}+3S_{-2,2,1}+4S_{1,-3,1}\nonumber\\
&&-4S_{1,-2,-2}-S_{1,-2,2}+16S_{1,1,-3}+34S_{1,1,3}+7S_{1,2,-2}+42S_{1,2,2}+34S_{1,3,1}\nonumber\\
&& +7S_{2,-2,1}+7S_{2,1,-2}-2\big(-21S_{2,1,2}-21S_{2,2,1}-17S_{3,1,1}+7S_{-2,1,1,1}\nonumber\\
&&+5S_{1,-2,1,1}+7S_{1,1,-2,1}+9S_{1,1,1,-2}+28(S_{1,1,1,2}+S_{1,1,2,1}+S_{1,2,1,1}+S_{2,1,1,1})\nonumber\\&&-20S_{1,1,1,1,1} \big)\big)\Big)\,.
\end{eqnarray}
Combining these contributions together results in \eqref{twisttwothreeloops}.
%%%%%%%%%%%%%%%%%%%%%%%%%%%%%%%%%%%%%%%%%%%%%
\subsection{Twist-three operators}\label{app:dimtwist3}

In analogy to the twist-two operators we choose to write the leading-order
Baxter function as
\begin{equation}
Q^{(0)}(u)=\sum_{k=0}^{\frac{M}{2}}\frac{b_{k,M/2}}{(k!)^2}\,(\tfrac{1}{2}+iu)_k\,(\tfrac{1}{2}-iu)_k\,.
\end{equation}
To obtain the expressions that contribute to the anomalous dimension we have to introduce
a regulator, i.e.~we analyze the derivatives of $Q^{(i)}(u)$ at $u=\pm i(\sfrac{1}{2}+\epsilon)$
and take the limit $\epsilon \to 0$. All sums now have the argument $\sfrac{M}{2}$.
\begin{eqnarray}
Q^{(0)}{}'(-\tfrac{i}{2})&=&i\sum_{k=1}^{\frac{M}{2}}\frac{b_{k,M/2}}{k}
			   =2\,i\,S_1\,,\\
Q^{(0)}{}''(-\tfrac{i}{2})&=&i^2\sum_{k=1}^{\frac{M}{2}} \frac{b_{k,M/2}}{k^2}
			    =4\,i^2\big(2S_{1,1}-S_{2}\big)\,,\\
Q^{(0)}{}'''(-\tfrac{i}{2})
&=&6\,i^3\sum_{k=1}^{\frac{M}{2}} \frac{b_{k,M}}{k}\,S_2(k-1)=24\,i^3\big(2\,S_{1,1,1}-S_{1,2}-S_{2,1}\big)\,.
\end{eqnarray}
One easily verifies that
\begin{equation}
Q^{(1)}(-\tfrac{i}{2})=c(M)\,,
\end{equation}
\begin{equation}
Q^{(1)}{}'(-\tfrac{i}{2})=2\,i\big(c(M)S_{1}-2S_{1,2}-2S_{2,1}-S_{3}\big)\,.
\end{equation}
According to \eqref{A} and \eqref{B} the building blocks of the anomalous dimension become
\begin{eqnarray}
A_{-}(M)&=&-8S_{3}\,. \\
B_{-}(M)&=&12S_{3}+8S_{1}S_{2}\,,
\end{eqnarray}
such that $\gamma^{(1)}$ is given by \eqref{gamma1 twist3}.

\medskip

\noindent
{\bf Three-loops.} The NNLO anomalous dimension requires the additional derivatives of $Q^{(0)}$
\begin{eqnarray}
Q^{(0)}{}^{(4)}(-\tfrac{i}{2})
=24\,i^4\sum_{k=1}^{\frac{M}{2}} \frac{b_{k,M/2}}{k^2}\,S_2(k-1)&=&
96\,i^4\big(\,2\,( 2\,S_{1,1,1,1}-S_{1,1,2}-S_{1,2,1}-S_{2,1,1})\nonumber\\
	&&+S_{2,2}+S_{3,1}\big)\,,
\end{eqnarray}
\begin{eqnarray}
Q^{(0)}{}^{(5)}(-\tfrac{i}{2})
&=&60\,i^5\sum_{k=1}^{\frac{M}{2}} \frac{b_{k,M/2}}{k}\,\big(S_4(k-1)-S_2(k-1)^2\big)=
960\,i^5\big(2(2S_{1,1,1,1,1}-S_{1,1,1,2}\nonumber\\
&&-S_{1,1,2,1}-S_{1,2,1,1}-S_{2,1,1,1})+S_{1,2,2}+S_{2,1,2}+S_{2,2,1}+S_{1,3,1}\big)\,.
\end{eqnarray}
Higher derivatives of the NLO Baxter function are given by
\begin{eqnarray}
Q^{(1)}{}''(-\tfrac{i}{2})&=&4\,i^2\big((2S_{1,1}-S_{2})c(M)-8(S_{1,1,2}+S_{1,2,1}+S_{2,1,1})\nonumber\\
&&+4S_{1,3}+10S_{3,1}+8S_{2,2}-3S_{4}\big)\,,\\
Q^{(1)}{}'''(-\tfrac{i}{2})&=&
24\,i^3\big((2S_{1,1,1}-S_{1,2}-S_{2,1})c(M)-12(S_{1,1,1,2}+S_{1,1,2,1}+S_{1,2,1,1}+S_{2,1,1,1})\nonumber\\
&&+2(5S_{1,1,3}+6S_{1,3,1}+S_{3,1,1}+6S_{1,2,2}+6S_{2,1,2}+6S_{2,2,1})-3S_{1,4}+S_{4,1}\nonumber\\
&&-7S_{2,3}-3S_{3,2}\big)\,.
\end{eqnarray}
The three-loop correction contributes with the terms
\begin{eqnarray}
Q^{(2)}(-\tfrac{i}{2})&=&c^{(2)}(M)\,,\nonumber\\
Q^{(2)}{}'(-\tfrac{i}{2})&=&2i\Big(c^{(2)}(M)S_1-c(M)(S_3+2S_{1,2}+2S_{2,1})+
 2\big(11S_{1,4}+19S_{4,1}-3S_{2,3}+5S_{3,2}  \big) \nonumber\\
&&+8\big(S_{1,2,2}+S_{2,1,2}+S_{2,2,1}-2S_{1,3,1}-4S_{3,1,1}\big)-15S_{5} \Big)\,.
\end{eqnarray}
Plugging all terms into the expansion formula for $\gamma^{(2)}$ and transforming to the
canonical basis one obtains \eqref{gamma2 twist3}.
%%%%%%%%%%%%%%%%%%%%%%%%%%%%%%%%%%%%%%%%%%%%%%%%%%%%%%%%%%%%%%%%%%%%%%
\section{A general method for solving an inhomogeneous Baxter equation}\label{app:Problems}
In this appendix we formulate a method for solving \textit{any} consistent inhomogeneous Baxter 
equation. For simplicity we confine ourselves to the case of twist-two operators, where the 
transfer matrix is unambiguously determined, but the generalization to more complicated cases 
(upon knowing the solution to the homogeneous equation) should be straightforward.\vspace{2mm}
\noindent
{\bf{Lemma 1.}}
A minimal polynomial solution to the inhomogeneous Baxter equation of the form
\begin{equation}\label{general inhomo}
\left(u+\tfrac{i}{2} \right)^2 Q^{(\ell)}(u+i)+\left(u-\tfrac{i}{2}\right)^2 Q^{(\ell)}(u-i)
-t^{(0)}_2(u)Q^{(\ell)}(u)=\sum_{k=0}^{M-1} \alpha_k (\tfrac{1}{2} \pm iu)_k \,,
\end{equation}
with the twist-two transfer matrix $t^{(0)}_2(u)=2 \, u^2 -(M^2+M+\tfrac{1}{2})$ and 
arbitrary coefficients $\alpha_k$ that are independent of the spectral parameter $u$ is given by
\begin{equation}
Q^{(\ell)}=\sum_{k=0}^{M}\frac{\beta_k \, R(k,M)}{k!}\,(\tfrac{1}{2} \pm iu)_k-\frac{\beta_M \,R(M,M) i^M}{M!} Q^{(0)}(u)\,
\end{equation}
with
\begin{equation}
\beta_k=\sum_{j=1}^{k}\frac{\alpha_{j-1}\,(j-1)!}{j^2 \, R(j,M)}\,, \quad R(k,M)=
\frac{(-1)^k\,\Gamma(k+M+1)}{(k!)^2\,\Gamma(M+1-k)}\,.
\end{equation}
and $Q^{(0)}(u)$ being the solution to the homogeneous equation with the coefficient 
in front of the highest power of u normalized to one.

The proof is easily obtained using the methods of \Appref{App: Mellin}.

\medskip

This lemma allows to write down the solution to the class $\III$ terms discussed in 
section \ref{Twist-two operators}, as well as to all other inhomogeneities considered 
in this paper. It must be noted, however, that this method does not lead necessarily 
to the simplest representation of the solution.

Below we will show how to bring the above-mentioned inhomogeneities of class $\III$ to a 
form, for which the Lemma 1 is directly applicable. The terms in question are a real 
combination of the following function
\begin{equation}\label{A(u)}
\left[\frac{1}{\left(u+\frac{i}{2}\right)^2}
+\frac{i\, \gamma^{(0)}}{2\left(u+\frac{i}{2}\right)}
\right]Q^{(0)}(u+i) -\frac{1}{(u+\frac{i}{2})^2} \, Q^{(0)}(u)=A(u)\,,
\end{equation}
Expanding around $u=-\frac{i}{2}$ allows one to check that $A(u)$ is a polynomial of 
degree $\deg A(u)=\deg Q^{(0)}(u)-1$. Moreover the real combination
$A(u)+A^{*}(u)=B(u)$ is a real polynomial of degree $\deg B(u)=(\deg Q^{(0)}(u)-2)$. 
In order to find the closed form of $A(u)$ in \eqref{A(u)} we note that
\begin{equation}
Q^{(0)}(u)=i(u+\tfrac{i}{2})M(M+1)\,{}_4 F_3\left(\left. \begin{array}{c}
					1-M, \ M+2,\ \frac{3}{2}-iu,\ 1 \\
				 	2,\  2 ,\ 2 \end{array}
					\right| 1\right)+1\,.
\end{equation}
The key fact that makes $A(u)$ a polynomial is that $\gamma^{(0)}$ can
be written in a similar fashion, since
\begin{equation}
S_1(M)=\frac{M(M+1)}{2}\,{}_4 F_3\left(\left. \begin{array}{c}
					1-M, \ M+2,\ 1,\ 1 \\
				 	2,\  2 ,\ 2 \end{array}
					\right| 1\right).
\end{equation}
This allows to write $A(u)$ as
\begin{eqnarray}
A(u)&=&\frac{\gamma^{(0)}}{2}M(M+1)\,{}_4 F_3\left(\left. \begin{array}{c}
					1-M, \ M+2,\ \frac{1}{2}+iu,\ 1 \\
				 	2,\  2 ,\ 2 \end{array}
					\right| 1\right) \nonumber\\
&&-\frac{iM(M+1)}{u+\tfrac{i}{2}}\sum_{k=0}^{M-1}\frac{(1-M)_k(M+2)_k}{(2)_k{}^3}
	\left((\tfrac{1}{2}+iu)_k+(\tfrac{3}{2}-iu)_k-2(1)_k \right)\,.
\end{eqnarray}
Subsequently, using the following identities
\begin{equation}\label{eq:poch to exp}
(x)_k=\sum_{n=0}^{k}(-1)^{k-n}s_1(k,n)(x)^n\,, \qquad (x)^k=\sum_{n=0}^{k}(-1)^{k-n}s_2(k,n)(x)_n\,,
\end{equation}
where $s_1$ and $s_2$ denote Stirling numbers of, respectively, first and second kind, one finds the final result
\begin{equation}\label{proper A(u)}
A(u)=\frac{\gamma^{(0)}}{2}M(M+1)\,{}_4 F_3\left(\left. \begin{array}{c}
					1-M, \ M+2,\ \frac{1}{2}+iu,\ 1 \\
				 	2,\  2 ,\ 2 \end{array}
					\right| 1\right)+
\sum_{k=0}^M\, a_k (\tfrac{1}{2}-iu)_k\,.
\end{equation}
The coefficients $a_k$ in the above formula are given by
\begin{eqnarray}\label{a_k}
a_k &= &\sum_{n=k+1}^{M}\frac{(-M)_{n+1}(M+1)_{n+1}}{(2)_n{}^3}\sum_{j=k+1}^{n}(-1)^{n-j}\,s_1(n,j)\times\nonumber\\
&&\sum_{m=k+1}^{j}\frac{\big((-1)^{m-1}-1\big)}{m!} \, (-j)_m \, (-1)^{m-k}\,s_2(m-1,k)\,.
\end{eqnarray}
The hypergeometric part of \eqref{proper A(u)} can be treated with the presented 
method of orthogonal polynomials. On the other hand, the explicit form of the $a_k$ 
coefficients in \eqref{a_k} allows for the application of lemma 1. The final result 
is given by \eqref{QIII_a} together with \eqref{QIII_b}.
%%%%%%%%%%%%%%%%%%%%%%%%%%%%%%%%%%%%%%%%%%%%%%%%%%%%%%%%%%%%%%%%%%%%%%

%%%%%%%%%%%%%%%%%%%%%%%%%%%%%%%%%%%%%%%%%%%%%%%%%%%%%%%%%%%%%%%%%%%%%%
\end{document}